\documentclass{epl}

\title{Vortex waves and the onset of turbulence in $^3$He-B}
\shorttitle{Vortex waves in $^3$He-B}

\author{K.L. Henderson\inst{1} \and C.F. Barenghi\inst{2}}
\institute{
  \inst{1} School of Mathematical Sciences, 
  University of the West of England,\\
  Bristol, BS16 1QY, UK\\
  \inst{2} School of Mathematics and Statistics, University of Newcastle,\\
  Newcastle upon Tyne, NE1 7RU, UK
}

\def\vn{{\bm v}^n}
\def\vs{{\bm v}^s}
\def\u{\bm u}
\def\bnabla{\bm{\nabla}}
\def\bcdot{\bm{\cdot}}
\def\nus{\nu^s}

\def\etal{\mbox{\it et al.\ }}

\def\bOmega{\bm \Omega}
\def\blambda{\bm \lambda}
\def\i{\mbox{\rm i$\,\!$}}
\def\etc{\mbox{\it etc}}
\renewcommand{\Im}{{\cal I}m}
\renewcommand{\Re}{{\cal R}e}

\begin{document}

\maketitle

\begin{abstract}
In a recent experiment Finne \etal discovered an intrinsic
condition for the onset of quantum turbulence in $^3$He-B, that
$q=\alpha/(1-\alpha')<1$, where $\alpha$ and $\alpha'$ are mutual
friction parameters. The authors argued that this condition
corresponds to Kelvin waves which are marginally damped, so
for $q>1$ Kelvin waves cannot grow in amplitude and trigger 
vortex reconnections and turbulence. 
By analysing both axisymmetric and non-axisymmetric modes of 
oscillations of a rotating superfluid, we confirm that in the 
long axial wavelength limit the simple condition $q=1$ is indeed the
crossover between damped and propagating Kelvin waves.
\end{abstract}

\section{Introduction}

A striking feature of superfluidity is the existence of quantised
vortex filaments. Quantised vortices are particularly interesting because,
unlike vortices in a classical turbulent fluid, they are discrete, 
stable topological defects.  Until recently the study of
quantised vorticity and quantum turbulence\cite{D} was mainly confined 
to $^4$He, but in the last 
few years there has been progress in other contexts\cite{BDV}, ranging from
atomic Bose--Einstein condensates to $^3$He, which is a fermionic superfluid. 

Our work is motivated 
by the recent discovery by Finne \etal\cite{Finne} of a sharp transition 
at the temperature $T=0.6~T_c$ between
two distinct hydrodynamical regimes in $^3$He-B, 
where $T_c$ is the critical temperature. 
In their experiment Finne 
\etal injected one or a few seeding vortex loops into rotating $^3$He-B.
Using a non-invasive NMR measurement technique, they measured the total
length of the quantised vortices as a function of time.
They found that if $T>0.6~T_c$ 
the initial loops expand and straighten into rectilinear vortices which
align along the axis of rotation. However, if
$T<0.6 T_c$, the seeding loops evolve into a turbulent vortex tangle.
The effect is independent of the fluid's velocity and the details of
the initial condition, which suggests the existence of a remarkable
intrinsic criterion for the onset of quantum turbulence. On the
contrary, the condition for the onset of turbulence in a classical
fluid, that the Reynolds number is large, is extrinsic, because the
Reynolds number depends on the geometry and the scale of the flow, not only
on properties of the fluid.

Finne \etal interpreted their experimental result in terms
of Kelvin waves on the vortex filaments - helical 
perturbations of the position of a vortex core away from the unperturbed 
straight shape (see Fig.~1).
They argued that the superfluid vorticity is determined by the
ratio $q$ of dissipative and inertial forces in the superfluid, in analogy 
with the classical Reynolds number,
which expresses the ratio of inertial and viscous
forces in an ordinary Navier--Stokes fluid. The quantity $q$ is defined as

\begin{equation}
q=\frac{\alpha}{(1-\alpha')},
\label{q}
\end{equation}

\noindent
where $\alpha$ and $\alpha'$ are known\cite{Bevan} mutual friction coefficients,
which determine the strength of the interaction between the superfluid
vortices and the normal fluid. Finne \etal noticed that, for 
a single superfluid vortex, it was predicted\cite{BDV} that $q=1$ is the 
crossover from
Kelvin waves which propagate ($q<1$) and  Kelvin waves which are overdamped
($q>1$). By performing enlightening numerical simulations, Finne \etal
confirmed that, if $q>1$ an injected vortex loop straightens and becomes
rectilinear, whereas if $q<1$  Kelvin waves  
propagate freely, as predicted. In the latter case, the amplitude of
the Kelvin wave grows,
driven by the local difference between
the normal fluid velocity and the velocity of the vortex line\cite{ABC}.
When the waves' amplitude becomes of the same order of the average 
intervortex spacing, the
vortices reconnect with each other, quickly forming a disordered
turbulent tangle. The value of $q$ depends on $T$, and the observed
temperature $0.6~T_c$ corresponds to $q=1.3$, which is indeed close to
unity.

The aim of our work is to reconsider the condition $q=1$ (which for
a single vortex is the crossover from free to overdamped
motion) for a large density of vortices, such as the vortex lattice in the
rotating helium experiment of Finne \etal. 
In this the context  the behaviour of the
superfluid is described by the Hall-Vinen equations\cite{HV}. Thus, rather
than individual vortex lines, we consider a continuum of vortex lines.
The advantage of this model is that it allows us to explore effects 
which the theory of a single vortex filament cannot describe, notably 
the presence of boundaries and the degrees of freedom represented by the 
coherent oscillatory motion of many vortices. Mathematically, this 
corresponds to the fact that for a single vortex, since the core radius is
fixed, only $m=1$ modes (sideway displacements of the vortex core)
are possible, whereas for a bundle of
vortices any azimuthal $m$ symmetry is permitted.
The simple question which we address is therefore the following:
given a superfluid in a cylindrical container of radius $a$ rotating at
constant angular velocity $\Omega$, thus forming a vortex lattice with
a large density of vortex lines, does the condition $q=1$ represent
the crossover from propagating Kelvin waves to overdamped Kelvin waves
for {\it all} possible modes of oscillations ?

\section{The governing equations and the basic state}

The equation of motion of the superfluid in a coordinate system rotating 
with angular velocity $\bm{\Omega}=\Omega {\bf{e}}_z$ may be written as
\begin{eqnarray}
 \frac{\partial \vs}{\partial t}
+ (\bm{\vs \cdot\nabla }) \vs
&=&  \bnabla \Psi +
2\vs\times\bOmega +\alpha\widehat{\blambda}\times[\blambda \times(\vs-\vn)]
\nonumber \\[-1.5ex]
&& \label{eq:vs}\\[-1.5ex]
\nonumber
&+&\alpha'\blambda \times(\vs-\vn)- \alpha\nus\widehat{\blambda}\times
(\blambda \bcdot \bnabla)\widehat{\blambda}
+\nus(1-\alpha')(\blambda \bcdot \bnabla)\widehat{\blambda}
\end{eqnarray}
where $\vs$ and $\vn$ are the superfluid and normal fluid velocities in 
the rotating frame, $\blambda=\bnabla \!\times \!\vs + 2\bOmega$, 
$\widehat{\blambda}=\blambda/|\blambda|$ is the unit vector in the
direction of $\blambda$ and $\Psi$ is a collection of scalar 
terms.  Given the high viscosity of $^3$He-B, we assume that the normal fluid
is in solid body rotation around the $z$-axis, thus in the rotating frame
$\vn=0$.  Eq.~(\ref{eq:vs})
must be solved under the condition that $\bnabla \!\bcdot \!\, \vs=0$.
%
%
%
%
%
The quantity 
$\nus=(\Gamma / 4\pi) \log (b_{_0}/a_{_0})$ is the vortex tension
parameter, $\Gamma$ is the quantum of circulation, $a_{_0}$ is the
vortex core radius and $b_{_0}=(|\blambda|/\Gamma)^{-1/2}$ is the
average distance between vortices. 
The unperturbed vortex lattice
corresponds to the basic state $\vs_0=\vn_0=0$, $\nabla\Psi_0=0$
for which $\blambda=2\Omega {\bf{e}}_z$, working in cylindrical
coordinates $(r,\phi,z)$.

\section{The equations of the perturbations}

We perturb the basic state by letting $\vs=\u=(u_r,u_\phi,u_z)$, 
$\Psi=\Psi_0+\psi$,
where $|\u|\ll 1$ and $|\psi|\ll 1$ are small perturbations. We 
assume normal modes of the form
$\exp(\i\sigma t+\i m\phi + \i kz)]$, where $m$ and $k$ are respectively
the azimuthal and axial wavenumbers and $\sigma$ is the growth rate.
The aim of our calculation is to determine the real and imaginary parts
of $\sigma$, namely $\Re(\sigma)$ and $\Im(\sigma)$.
The resulting linearised equations for the perturbations are

\begin{eqnarray}
(\i\sigma +\alpha \eta) u_r
- \eta(1-\alpha')u_{\phi}
&=&\frac{d\psi}{dr}
-\frac{mk\nus}{r}(1-\alpha') u_z
-\i k\nus\alpha \frac{du_z}{dr}=0, 
\label{eq:ur} \\
\eta(1-\alpha') u_r
+(\i\sigma +\alpha \eta) u_{\phi}
&=&\frac{\i m}{r}\psi 
+\frac{\alpha m k \nus}{r} u_z
-\i k\nus(1-\alpha')\frac{du_z}{dr}=0,
\label{eq:uphi} \\[1.3ex]
\i\sigma u_z
&=&\i k\psi,
\label{eq:uz}\\
\frac{1}{r} \frac{d}{dr}(r\frac{du_r}{dr})+\frac{\i m}{r}u_{\phi}
+\i k u_z&=&0,
\label{eq:cont}
\end{eqnarray}
\noindent
where $\eta=2\Omega+\nus k^2$.
Eliminating $u_r$, $u_{\phi}$ and $\psi$ we obtain the following differential 
equation for $u_z$:
\begin{equation}
\frac{1}{r} \frac{d}{dr}\left(r\frac{du_z}{dr}\right)-\frac{m^2}{r^2}u_z+\beta^2 u_z=0,
\label{eq:bessel}
\end{equation}
\noindent
where
\begin{equation}
\beta^2=\frac{-k^2[(\i\sigma+\alpha \eta)^2 +(1-\alpha')^2 \eta^2]}
{[(\i\sigma+\alpha \eta)(\i\sigma+k^2 \nus \alpha)
+(1-\alpha')^2\nus k^2 \eta]},
\label{eq:beta}
\end{equation}
\noindent
The solution of Eq.~(\ref{eq:bessel}) which is regular as $r\to 0$ is
the Bessel function of the first kind of order $m$, $J_m(\beta r)$.
To determine $\beta$ we enforce the boundary condition $u_r=0$
on the wall of the container $r=a$, which yields the secular equation
\begin{equation}
\frac{(ka)^2}{(\beta a)}\frac{J'_m(\beta a)}{J_m(\beta a)} 
[(\i\sigma +\alpha \eta)^2 +(1-\alpha')^2\eta^2]
+2m\Omega \sigma (1-\alpha')=0,
\label{eq:eigen}
\end{equation}
\noindent
where the prime denotes the derivative of the Bessel function with respect to
its argument.

\section{Limiting cases}

The dispersion law which we have obtained, represented by
Eq.'s~(\ref{eq:beta},\ref{eq:eigen}), has two limiting cases 
which have already appeared in
the literature. Firstly, if we set $\alpha=\alpha'=\nus=0$, we recover
Chandrasekhar's classical result for the modes of vibrations of a 
rotating column of liquid\cite{Chandra}: $\sigma$ is determined by
\begin{equation}
\beta^2=k^2(\frac{4\Omega^2}{\sigma^2}-1),
\label{eq:chandra1}
\end{equation}
\noindent
where the values of $\beta$ are given by the roots of
\begin{equation}
\sigma \beta a J'_m(\beta a) + 2m\Omega J_m(\beta a)=0.
\label{eq:chandra2}
\end{equation}
\noindent
Secondly, if we set $\alpha=\alpha'=m=0$, Eq.~(\ref{eq:eigen}) reduces to
$J'_0(\beta a)=0$. Let $\xi_j$ be the $j^{th}$ zero of $J_1(\xi)=J'_0(\xi)$
(that is $\xi_1=3.83171,$ $\xi_2=7.01559$ \etc); 
then we have
\begin{equation}
\sigma^2=\frac{(k^2 a^2 \eta^2 +\nus \eta k^2 \xi^2_j)}
{(\xi^2_j+k^2 a^2)},
\label{eq:hall1}
\end{equation}
\noindent
If the wavelength of the perturbations is larger than the radius of the
cylinder we may neglect edge effects, $(ka)^2\gg\xi_j^2$, and we recover
Hall's result\cite{Hall} for the axisymmetric oscillations of a 
rotating superfluid:
\begin{equation}
\sigma^2 \approx \eta^2=2\Omega+\nus k^2.
\label{eq:hall2}
\end{equation}
\noindent
Eq.~(\ref{eq:hall2}) generalises to a vortex lattice the dispersion
relation $\sigma=\nus k^2$ of an individual vortex line\cite{BDV}.

\section{The axisymmetric case}

Eq.'s~(\ref{eq:beta},\ref{eq:eigen}) simplify in the axisymmetric
case, $m=0$. We have again $J'_0(\beta a)=0$ for which $\beta a=\xi_j$
and Eq.~(\ref{eq:eigen}) becomes
\begin{equation}
\frac{\xi^2_j}{(ka)^2}=
-\frac{[(\i\sigma +\alpha \eta)^2+(1-\alpha')^2\eta^2]}
{[(\i\sigma+\alpha \eta)(\i\sigma+\alpha \nus k^2)
+(1-\alpha')^2\nus \eta k^2]}.
\label{eq:axi1}
\end{equation}
This quadratic equation has solutions
\begin{eqnarray}
((ak)^2\!\!&+\!&\xi_j^2)\sigma =\i\alpha(2\Omega (ak)^2+\Omega \xi_j^2+
\nus k^2((ak)^2+\xi_j^2))
\nonumber \\[-1.0ex]
&& \label{eq:sigsol}\\[-1.0ex]
\nonumber
&\pm& \sqrt{\left\{(1-\alpha')^2k^2((ak)^2+\xi_j^2)(2\Omega+\nus k^2)
(2\Omega a^2+\nus((ak)^2+\xi_j^2)))
-\alpha^2\Omega^2\xi_j^4\right\}}
\end{eqnarray}
This result is identical to that derived by Glaberson \etal~\cite{glab:74} 
who expanded in normal modes of the form $\exp(\i k_1x+\i k_2y+\i kz+\i\sigma t)$
where $\xi_j^2/a^2=k_1^2+k_2^2$. Expanding in normal modes of this form means that the boundary condition at $r=a$ is not enforced.

From Eq.~(\ref{eq:sigsol}) it can be seen that $\Im(\sigma)$ is non-negative,
so the system is always stable to infinitesimal disturbances.  
If we neglect perturbations in the non-axial direction 
$({\rm setting} \ \beta=\xi=k_1=k_2=0)$ we obtain
\begin{equation}
\sigma=\i\alpha (2\Omega +\nus k^2) \pm (1-\alpha') (2\Omega +\nus k^2),
\label{eq:axi2}
\end{equation}
as quoted by Hall~\cite{Hall}.  This solution is illustrated in              
Fig.~\ref{fig:m0} by a dashed line and satisfies
$|\Im(\sigma)/\Re(\sigma)|=q$.  
However, if we do not neglect perturbations in the non-axial direction,
Eq.~(\ref{eq:sigsol}) has an infinite number of solutions, according to the
value of $\xi_j$ considered.  The least 
stable mode is the one for which $\Im(\sigma)$ is minimum and is
the mode which we would expect to be observed providing the initial state is
sufficiently random.  $\Im(\sigma)$ decreases monotonically as 
$k\rightarrow 0$ and $\xi_j\rightarrow\infty$.
For small $k$ the term inside the square root of Eq.~(\ref{eq:sigsol})
is negative, resulting in $\Re(\sigma)$ being zero and the mode decays 
monotonically with time.  As $k$ increases,
the term inside the square root is positive and the mode displays
oscillatory decay with time.
The solution which has the smallest
decay rate is
\begin{equation}
\sigma =\i\alpha(\Omega +\nus k^2)
\pm \sqrt{\left\{(1-\alpha')^2\nus k^2(2\Omega+\nus k^2)
-\alpha^2\Omega^2\right\}}
\end{equation}
which corresponds to $\xi_j \rightarrow \infty$.
Plots of the decay rate, $-\Im(\sigma)$ and the ratio of the decay 
rate of the wave and its angular frequency, $|\Im(\sigma)/\Re(\sigma)|$ can 
be found in Fig.~\ref{fig:m0}, where the mutual friction parameters 
correspond to $T=0.4T_c$, for which $\alpha=0.1125$ and $\alpha'=0.1$, 
and we have taken $\nus/a^2=0.001$ and $\Omega=1$.
For large $(ak)$ we find
\begin{equation}
\left|\Im(\sigma)/\Re(\sigma)\right|\rightarrow \frac{\alpha}{(1-\alpha')}=q.
\end{equation}
The minimum permitted value
of $k$ will be governed by the height, $h$ of the apparatus by the relation
$k_{\rm min}=2\pi/h$, so provided the aspect ratio $h/a$ is small 
enough we find agreement with the argument of Finne \etal~\cite{Finne}.

\section{Non-axisymmetric case}

In order to consider the non-axisymmetric modes we must solve the coupled
equations (\ref{eq:beta},\ref{eq:eigen}).  For given temperature, $T$, 
wavenumbers $m$, $ak$ and parameters $\nus/a^2$, $\Omega$
Eq.'s (\ref{eq:beta},\ref{eq:eigen}) were solved using
NAG routine C05NDF on the real and imaginary parts of $\sigma$ and 
$(a\beta)^2$.  As for the axisymmetric case, there will be an infinite number 
of solutions for each mode, $m$ considered.

In Fig.~\ref{fig:m1} we illustrate some of the solutions computed for the 
$m=1$ mode.  The quantities $-\Im(\sigma)$ and $|\Im(\sigma)/\Re(\sigma)|$ 
are plotted against $ak$ 
using the same parameters as for Fig.~\ref{fig:m0}. The dotted
line represents the least stable axisymmetric mode.  We can see that
the least stable $m=1$ mode is bound by the least stable axisymmetric
mode and that, as for the $m=0$ case, all the computed solutions show
$|\Im(\sigma)/\Re(\sigma)|\rightarrow q$ at large values of $ak$.
 
In Fig.~\ref{fig:allm} we plot $-\Im(\sigma)$ and 
$|\Im(\sigma)/\Re(\sigma)|$ against $ak$ of the least stable computed modes
for $m=0,2,4,8,11,\infty$ using the same parameters as for Fig.~\ref{fig:m0}. 
The arrow shows the direction of increasing $m$. 
From the values of $\Im(\sigma)$ it can be seen that all the modes are 
stable and that the least stable mode occurs as $m\rightarrow \infty$.  
For all the modes we find that $|\Im(\sigma)/\Re(\sigma)|\rightarrow q$
for large values of $ak$ and using the same argument put forward for the
axisymmetric case, we conclude that we find agreement with the argument of Finne
provided the aspect ratio of the apparatus is small enough.

As $m\rightarrow\infty$ we are able to make a simplification to~Eq.~(\ref{eq:eigen}).
From Abromavitch \& Stegun~\cite{abromstegun} we find 
\begin{equation}
 J_m(z)\approx \frac{1}{\sqrt{2\pi m}}\left(\frac{ez}{2m}\right)^m
\hspace{8mm} {\rm as } \ \ \ m\rightarrow \infty,
\end{equation}
so, in this limit, we may take
\begin{equation}
 \frac{J'_m(a\beta)}{(a\beta)J_m(a\beta)}\approx\frac{m}{(a\beta)^2}.
\end{equation}
Substituting into Eq.~(\ref{eq:eigen}) we find that the resulting equation 
is independent of $m$.
This equation may be solved analytically and yields two solutions for $\sigma$,
namely $\sigma=(1-\alpha')\nus k^2 +\i\alpha\nus k^2$ and 
$\sigma=-(1-\alpha')\eta +\i\alpha\eta$.  The first solution will be the least
stable and this is the one that has been plotted in Fig.~\ref{fig:allm}.
Note that both of these solutions are
such that $|\Im(\sigma)/\Re(\sigma)|=q$. 

\section{Conclusion}

In conclusion, our analysis shows that in the case of
both axisymmetric ($m=0$) and non-axisymmetric ($m \ne 0$) perturbations of
a vortex lattice, provided that the aspect ratio
is small enough ($h/a\gg 1$), Kelvin waves propagate if $q<1$ and
are damped if $q>1$. This result confirms the argument put forward by
Finne \etal.  Finally, the exact dispersion relation which we have found
(Eq.~\ref{eq:beta} and \ref{eq:eigen}) can be used to study with
precision the spectrum of Kelvin waves on a rotating vortex lattice for
any height and radius of experimental apparatus.

\begin{figure}
\begin{center}
\includegraphics[width=3.0cm]{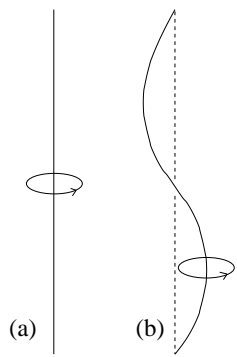}
\caption{(a): unperturbed vortex; (b): Kelvin wave.}
\label{fig:1}
\end{center}
\end{figure}

\begin{figure}
\begin{center}
\includegraphics[width=12.0cm]{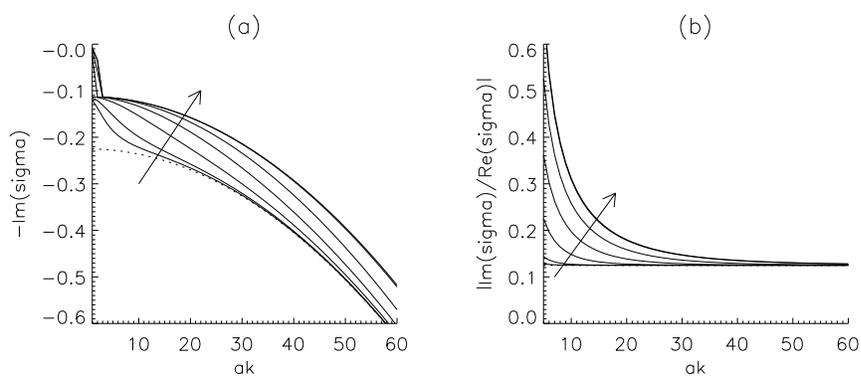}
\caption{Plots of (a) the decay rate, $\Im(\sigma)$ 
(b) $|\Im(\sigma)/\Re(\sigma)|$ of the axisymmetric mode $(m=0)$
against $ak$ for various values of $\xi_j$.
The arrows show the direction of increasing $\xi_j$, where 
$j=1,2,5,10,20,100,\infty$. The dashed line shows the result for $\xi_j=0$.}
\label{fig:m0}
\end{center}
\end{figure}

\begin{figure}
\begin{center}
\includegraphics[width=12.0cm]{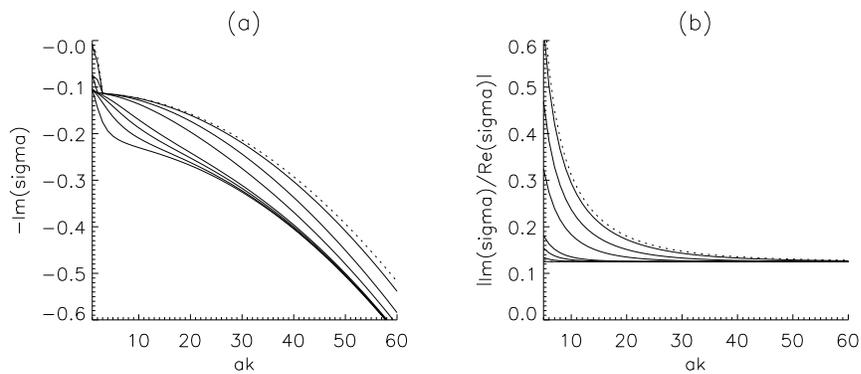}
\caption{Plots of (a) the decay rate, $\Im(\sigma)$ 
(b) $|\Im(\sigma)/\Re(\sigma)|$ against $ak$ of the $m=1$ mode.
The dotted line represents the least stable axisymmetric mode.}
\label{fig:m1}
\end{center}
\end{figure}

\begin{figure}
\begin{center}
\includegraphics[width=12.0cm]{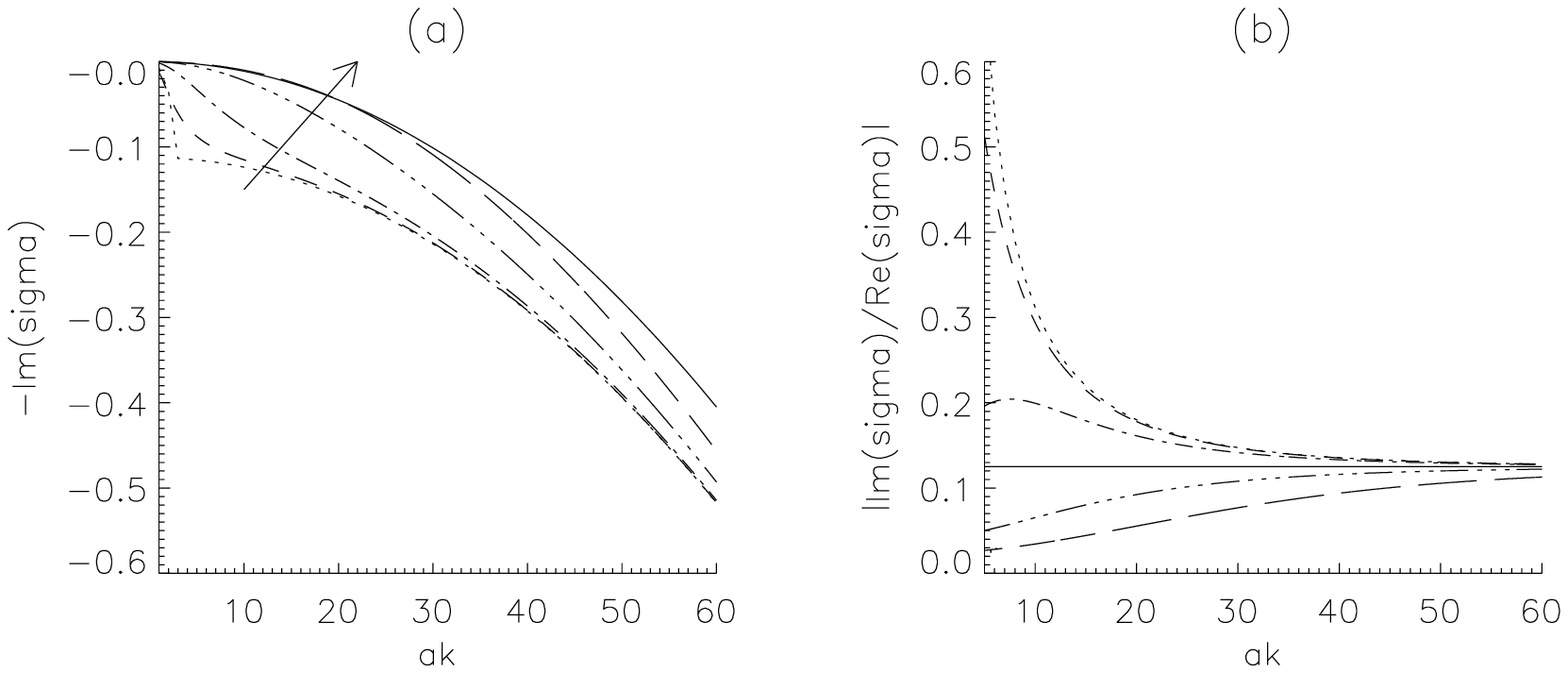}
\caption{Plots of (a) the decay rate, $\Im(\sigma)$ 
(b) $|\Im(\sigma)/\Re(\sigma)|$ against $ak$ for the $m=2,4,8,11$
mode. The dotted line represents the least stable axisymmetric mode.
The solid line represents the $m=\infty$ mode.  The arrow shows the 
direction of increasing $m$.} 
\label{fig:allm}
\end{center}
\end{figure}


\section{Acknowledgements}
C.F.B. is grateful to M. Krusius and members of his group for the
kind hospitality and many useful discussions.

\end{document}